\documentclass{llncs}

\title{Integrating splice-isoform expression into genome-scale models characterizes\\ breast cancer metabolism}
  \author{Claudio Angione}
  \institute{Department of Computer Science and Information Systems\\Teesside University\\ Middlesbrough, UK}
\usepackage{multirow}
\usepackage{makeidx}  % allows for indexgeneration

\usepackage{amssymb,amsfonts,amsmath}
\usepackage{graphicx}
\usepackage{tabularx}
\usepackage{algorithm}
\usepackage{algorithmicx}

\usepackage{algpseudocode}
\usepackage{epsfig}
\usepackage{verbatim} 
\usepackage{epstopdf}
\usepackage{hyperref}

\usepackage{xr}
\graphicspath{{images/}}

\begin{document}

\maketitle              % typeset the title of the contribution

This is a pre-copyedited, author-produced version of an article accepted for publication in Bioinformatics following peer review. The version of record ``C. Angione, "Integrating splice-isoform expression into genome-scale models characterizes breast cancer metabolism", Bioinformatics, btx562, 34(3):494-501, 2018'' is available online at: \url{https://academic.oup.com/bioinformatics/article-abstract/34/3/494/4107935}.

\abstract{\textbf{Motivation:} Despite being often perceived as the main contributors to cell fate and physiology, genes alone cannot predict cellular phenotype. During the process of gene expression, $95\%$ of human genes can code for multiple proteins due to alternative splicing.  While most splice variants of a gene carry the same function, variants within some key genes can have remarkably different roles.
To bridge the gap between genotype and phenotype, condition- and tissue-specific models of metabolism have been constructed. However, current metabolic models only include information at the gene level. Consequently, as recently acknowledged by the scientific community, common situations where changes in splice-isoform expression levels alter the metabolic outcome cannot be modeled. \\
\textbf{Results:} We here propose GEMsplice, the first method for the incorporation of splice-isoform expression data into genome-scale metabolic models. Using GEMsplice, we make full use of RNA-Seq quantitative expression profiles to predict, for the first time, the effects of splice isoform-level changes in the metabolism of $1455$ patients with $31$ different breast cancer types. We validate GEMsplice by generating cancer-versus-normal predictions on metabolic pathways, and by comparing with gene-level approaches and available literature on pathways affected by breast cancer.
GEMsplice is freely available for academic use at \url{https://github.com/GEMsplice/GEMsplice_code}.
Compared to state-of-the-art methods, we anticipate that GEMsplice will enable for the first time computational analyses at transcript level with splice-isoform resolution.
}

\section{Introduction}

The increasing availability of multi-omic datasets has rapidly moved the focus of many research efforts from data collection to finding effective methods for data interpretation and analysis. In terms of biomedical results, this step is therefore currently considered more limiting than data collection itself \cite{stephens2015big}.

Genes and their expression have been the subject of the vast majority of research studies in computational biology and bioinformatics. However, it is the interaction of genes, proteins, reactions, and metabolites (different \textit{omics}) that shapes the behavior of a cell. 
When analyzing biological models, the classical pathway-based perspective has been replaced, in the last 20 years, by a network-based approach, therefore bypassing parametrization and need for kinetic data, and leading to the generation of genome-scale metabolic models.  
These models include thousands of biochemical reactions, often  the full set of reactions known for a given organism, allowing for prediction of cell phenotype. 
In this regard, metabolism is the only biological system that can be fully modeled at genome-scale, and also the closest to the phenotype.

Metabolism is now considered as a driver, rather than a marker, of cancer onset and proliferation.  In breast cancer, metabolic heterogeneity is one of the causes of poor clinical outcome: although being classified as a single disease, more than $20$ types of breast cancer exist. A genome-scale analysis of cancer metabolism captures many effects that could not be identified using standard data and gene expression analysis.  Furthermore, a wide set of tools for the incorporation of cancer omic data into these models have been developed, making them suitable for integrating and interpreting the great amount of data that is being gathered on cancer metabolic alterations \cite{qi2017inference}.

As mentioned above, reduced cost for data gathering and analysis has recently yielded a rapid increase in the amount of available omics data. RNA-Seq is now widely used to produce high-throughput data in more detail compared to microarrays, with a more accurate estimation of transcript levels. As a result, splice isoform expression levels are now becoming available in cancer studies, where the same gene can code for multiple proteins due to alternative RNA splicing before translation. Alterations of specific splice-isoform expression in some genes can constitute a biomarker for cancer metabolism. In humans, alternative splicing affects $95\%$ of multiexon genes \cite{pan2008deep}.

However, to date, functional information included in most metabolic models refers to genes or proteins only. The idea exploited here is that splice isoform data can be readily integrated and used in conjunction with annotated genome-scale models in order to constrain and refine existing models.  Although Recon1 \cite{duarte2007global} was the only human metabolic model that introduced isoform-level annotations for some genes through bibliographic research, it adopted a custom annotation for isoforms, and therefore did not allow any mapping to known identifiers from public databases. For these reasons, such annotations have been subsequently lost and then simply ignored, mapping transcriptomic data with gene-level resolution only \cite{ryu2015reconstruction}. Consequently, expression data at the splice-isoform level has been neglected or simply averaged within the same gene to approximate the expression at the gene level.

Nevertheless, especially in human metabolic models, the incorporation of splice isoforms is key to understanding complex diseases like cancer. In this regard, pyruvate kinase (PK) is a striking example. In fact, the switch to the second isoform of pyruvate kinase (PKM2) is considered essential for cancer growth \cite{wong2015pkm2}. Conversely, the switch from PKM2 to the main pyruvate kinase isoform (PKM1) is able to reverse the Warburg effect and could therefore constitute a therapeutic target. In general, these crucial isoform-level events, e.g.\ the switch to minor isoforms, which get overexpressed compared to the major isoform of a gene, cannot be captured by current metabolic models, as highlighted in a number of recent reviews \cite{pfau2016towards,yizhak2015modeling,ryu2015reconstruction,geng2017silico}.

\begin{figure}
	\centering
	\includegraphics[width=\columnwidth]{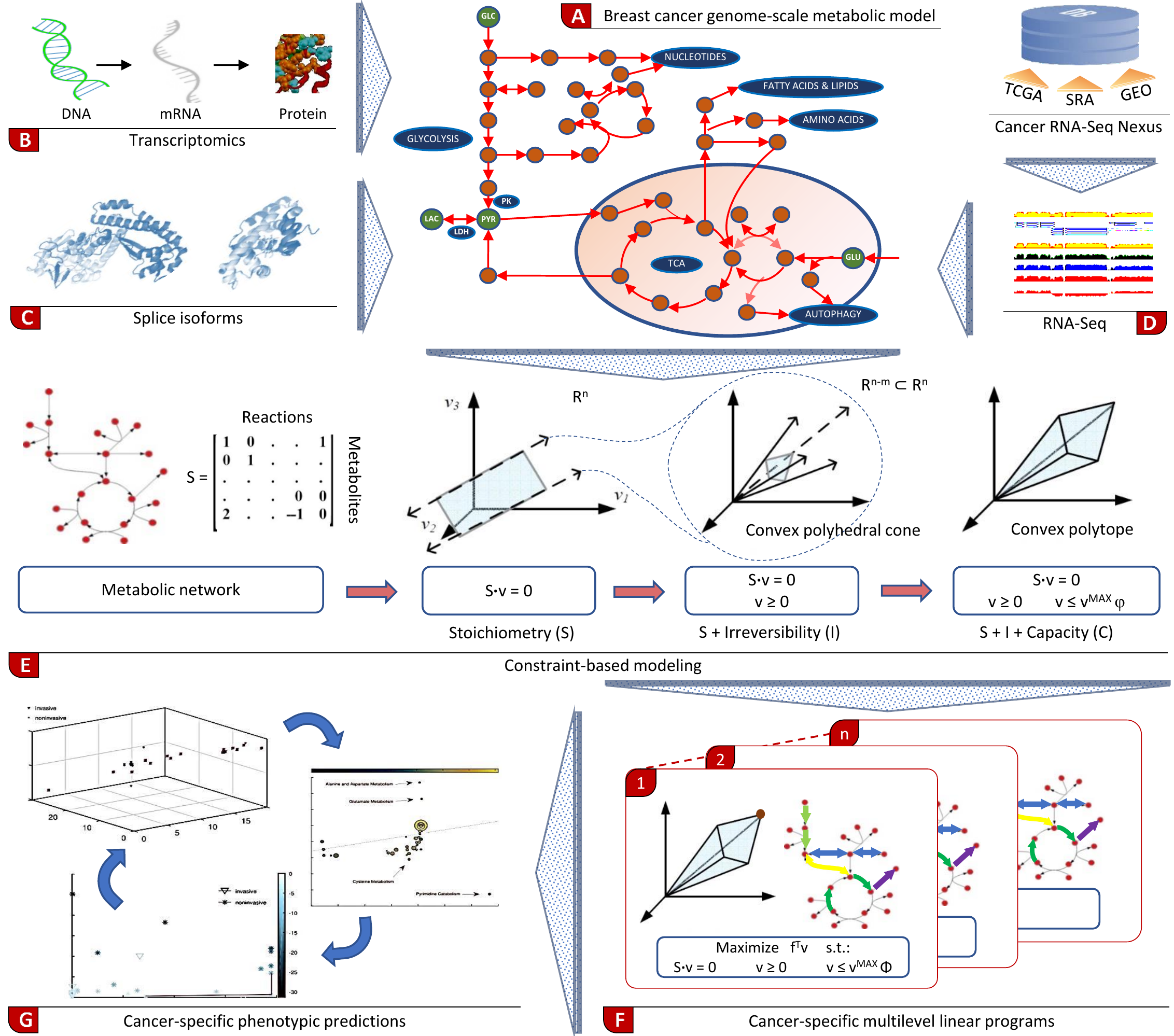}
	\linespread{1} %force single line spacing
	\vspace{-0.3cm}
	\caption{{\bf GEMsplice incorporates RNA-Seq data into genome-scale metabolic models at the splice-isoform level.} Starting from a metabolic model of breast cancer metabolism \cite{jerby2012metabolic} (A), gene expression (B) and transcript level information (C) are incorporated into the model. As a result, for the first time, we can exploit the full potential of next-generation sequencing in the context of genome-scale metabolic reconstructions.  A set of phenotype-specific RNA-Seq transcript expression levels in a variety of breast cancer types and stages from the Cancer RNA-Seq Nexus dataset \cite{li2016cancer}, including data from TCGA, GEO and SRA (D), are then mapped onto the model using constraint-based modeling (E). Cancer-specific metabolic models are finally generated and investigated using multilevel linear programming (F), leading to phenotype prediction for different types of breast cancer (G).}
	\label{fig:pipeline_full}
\end{figure}

Here we propose GEMsplice (genome-scale metabolic modeling with splice-isoform integration), the first method for incorporating RNA-Seq data at the splice-isoform level into a metabolic model.  The GEMsplice pipeline exploits, for the first time, the full potential of the next-generation sequencing technology in the context of genome-scale metabolic reconstructions. As a result, it enables more accurate predictions of human metabolic behavior and cancer metabolism. We show that GEMsplice compares favorably to existing tools for integration of transcriptomics data at the gene level only. We also validate GEMsplice by building breast cancer-versus-normal genome-scale models and by comparing predictions with available results on metabolic pathways affected by breast cancer. 
The full GEMsplice pipeline is illustrated in Figure \ref{fig:pipeline_full}. GEMsplice is made freely available in Matlab/Octave at \url{https://github.com/GEMsplice/GEMsplice_code}, and is fully compatible with the COBRA 2.0 Toolbox \cite{schellenberger2011quantitative}. 
%
%
%
%
%\enlargethispage{12pt}
%
\subsection*{Beyond the Warburg effect: metabolism is a key player in cancer formation and progression}
Known molecular markers of cancer can be associated with two main classes: (i) cell proliferation and (ii) tissue remodeling \cite{markert2015mathematical}. Cancer cells show  higher rate of proliferation than normal cells in the tissue from which they originated. 
The main goal of metabolism in cancer cells is to keep cell viability and ensure new biomass production by acquiring nutrients from an environment where nutrients are often scarce.
To this end, cancer cells exhibit a modified metabolism and an increased need for proteins, energy, nucleotides and lipids.
The two main nutrients used by cancer cells are glucose and glutamine; both support cell growth and are used to build carbon intermediates, which are employed in a number of processes that build macromolecules.

The first difference between normal and cancer metabolism was observed in the 1920s by Otto Warburg \cite{warburg1927metabolism}. Normal cells take up glucose and perform glycolysis to obtain pyruvate, which is then transferred to the mitochondrion and used by oxidative phosphorylation (OXPHOS) in the TCA cycle along with oxygen to efficiently produce ATP. Conversely, proliferating cancer cells require higher amount of glucose, but after obtaining pyruvate they preferentially use it to secrete lactate in the cytoplasm. Strikingly, this rather inefficient energy production pipeline is used also when the cell is exposed to oxygen. Preferential use of glycolysis allows faster proliferation but maximizes the secretion of lactic acid, which damages surrounding cells. This behavior, called \textit{Warburg effect} or \textit{aerobic glycolysis}, facilitates proliferation and migration of cancer cells.

However, a misinterpretation of this phenomenon by Warburg himself led to the enduring misconception that cancer cells do not use the TCA cycle to produce ATP. Indeed, it must be noted that the vast majority of cancer cells still use the mitochondrion and its TCA cycle to produce a fraction of the required ATP. In fact, in contrast to quiescent cells, glycolysis does not fuel directly the TCA cycle, but is essentially decoupled from it \cite{richardson2008central}. The advantage of converting excess pyruvate into lactate instead of transferring it into the mitochondrion is that high glycolytic activity can continue without leading to excessive flux through the electron transport system and consequent overproduction of reactive oxygen species (ROS), or excess ATP and NADH generation, which would in turn repress glycolysis. As a result of this decoupling, useful glycolytic intermediates (e.g., precursors of serine biosynthesis) can be generated at higher rates. 

Pyruvate kinase (PK), the last step in the glycolytic pathway, is responsible for keeping the balance between the production of pyruvate for the mitochondrion (high PK activity) and the production of glycolytic intermediates for biosynthesis (low PK activity). This is achieved through the preferential expression of PKM1 or PKM2, two splice isoforms of PK whose activity respectively supports pyruvate or glycolytic intermediates for biosynthetic processes. In cancer phenotypes, being mostly controlled by the isoform PKM2, PK and its ATP product are independent of oxygen, therefore enabling ATP generation and growth during hypoxia  (unlike mitochondrial respiration, which needs oxygen).

Glucose is not the only key nutrient for cancer cells. Proliferating cells can consume up to ten times more glutamine than any other amino acid. Glutamine is used as a carbon source for fatty acid synthesis, in biosynthetic pathways (especially the biosynthesis of nucleotides), and for TCA cycle intermediates (although as a secondary source, less than glucose).
While in normal cells glucose/glutamine intake depends mainly on extracellular stimuli, mutations undergone by cancer cells confer the ability to proliferate with a high degree of independence, by constantly importing amino acids and glucose from the extracellular environment \cite{yang2017glutaminolysis}. 
Taken together, these studies show that metabolism, once believed to be mainly a passive indicator of the state of a cell, is now widely recognized as a key player in cancer formation and progression.

\pagebreak
	\section{Methods}

	\enlargethispage{6pt}

	\subsection*{Incorporating RNA-Seq data into a breast cancer model at the splice-isoform level}
	To model splice-isoform data in the breast cancer metabolic model, $31$ RNA-Seq expression profiles of breast cancer were considered from Cancer RNA-Seq Nexus \cite{li2016cancer}, which provides phenotype-specific transcript expression levels in a variety of cancer types and stages. In Cancer RNA-Seq Nexus, data were processed as follows. For GEO/SRA samples, Bowtie \cite{langmead2009ultrafast} was used to align RNA-Seq reads to the reference human transcriptome from GENCODE \cite{harrow2012gencode}; eXpress \cite{roberts2013streaming} was then used to obtain FPKM isoform abundances. For TCGA samples, starting from Level 3 RNA-Seq v2 ``tau values'' quantified through RSEM \cite{li2011rsem}, TPM values were obtained through multiplication by $10^6$.

	The expression values, at this point, are measured in FPKM for GEO/SRA, and in TPM for TCGA samples. FPKMs were then converted to TPMs using the following formula for all the expression values $j$ in each sample $i$:
	$\mbox{TPM}_{ij} = 10^6 \frac{\mbox{FPKM}_{ij}}{\lambda_i \sum_j{\mbox{FPKM}_{ij}}}$, where $\sum_j{\mbox{FPKM}_{ij}}$ is the sum of the FPKM values of all the transcripts in the $i$th sample, and $\lambda_i$ a sample-specific scaling factor accounting for the fact that the number of transcripts varies across the different samples, $\lambda_i = $ (number of transcripts in the largest sample)$/$(number of transcripts in sample $i$).
	We remark that TPM (and not FPKM) expression values were used as a starting point to generate the metabolic outcome across samples because they are proportional to the abundance and independent of the mean expressed transcript length, therefore making them more suitable to comparisons across samples \cite{li2011rsem}.

	Transcript annotations in Cancer RNA-Seq Nexus consist of Ensembl and UCSC IDs. These were converted to RefSeq annotations using BioMart \cite{smedley2015biomart}. 
	However, the breast cancer model is annotated only with gene-level Entrez IDs. To expand these and generate transcript-level IDs with splice-isoform resolution,  RefSeq IDs for $141$ transcripts were retrieved from the SBML source of Recon1 using a custom script, as they were not included in the final Matlab version of the model (see File S1).
	These newly generated transcript-level RefSeq IDs and the gene-level default Entrez IDs of the remaining genes in the model were used to associate each transcript in the model with the corresponding expression value in the $31$ RNA-Seq cancer profiles.
	The profiles were finally mapped onto the breast cancer metabolic model (see Figure \ref{fig:pipeline_full}, and the following \textit{Methods} subsection for details on how the mapping was achieved).

	\subsection*{Phenotype predictions from RNA-Seq data}
	
	Breast cancer is a multifactorial disease and, as a result, breast cancer cells are intrinsically multi-target. To model this behavior and generate cancer-specific models from Cancer RNA-Seq Nexus, an extended flux balance analysis (FBA) framework was used, coupled with a multi-omic integration method and multi-level linear programming. Several approaches have been proposed and reviewed for the integration of gene expression data into FBA models \cite{machado2014systematic,vijayakumar2017seeing}. Each reaction in a FBA model is controlled by an associated gene set, defined through AND/OR Boolean operators between genes. In this work, METRADE \cite{angione2015predictive} was used to constrain the upper- and lower- limits of each reaction as a  function of the expression level of the gene set controlling the reaction.

	Multiple cellular targets were modeled as a trilevel linear program:
	\begin{equation}
	\begin{aligned}
	& {\text{max}} 
	& & h^\intercal v \\
	&  {\text{such that}}
	& & \text{max } g^\intercal v \\
	& & & {\text{such that}}
	& \text{max } f^\intercal v, \quad  Sv = 0,\\
	& & & & v^{\text{min}} \varphi(\Theta) \leq v \leq v^{\text{max}} \varphi(\Theta), % the \right. delimits \left| to have a big vertical bar for the "`such that"'
	\end{aligned}
	\label{eq:trilevel}
	\end{equation}
	where $S$ is the stoichiometry matrix of the metabolic reactions in the cell, $v$ is the vector of reaction flux rates, while $f$, $g$, $h$ are Boolean vectors of weights selecting the three reactions in $v$ whose flux rate will be considered as objective. Lower- and upper-limits for the flux rates in $v$ in the unconstrained model are given by the vectors $v^{\text{min}}$ and $v^{\text{max}}$. The vector $\Theta$ represents the gene set expression of the reactions associated with the fluxes in $v$. The expression level $\Theta$ of a gene set is defined from the expression levels $\theta(g)$ of its genes. According to the type of gene set, we define $\Theta(g) = \theta(g)$ for single genes, $\Theta(g_1 \wedge g_2) = \min\{ \theta(g_1), \theta(g_2)\}$ for enzymatic complexes, and $\Theta(g_1 \vee g_2) = \max\{ \theta(g_1), \theta(g_2)\}$ for isozymes.
	%
	%\begin{equation}
	%\begin{aligned}
	%\Theta(g) &= \theta(g)    & \mbox{(single gene)}, \\
	%\Theta(g_1 \wedge g_2) &= \min\{ \theta(g_1), \theta(g_2)\}     & \mbox{(enzymatic complex)}, \\
	%\Theta(g_1 \vee g_2) &= \max\{ \theta(g_1), \theta(g_2)\}   & \mbox{(isozyme)}.    
	%\end{aligned}
	%\label{eq:gene_genesets}
	%\end{equation}
	%
	These rules were applied recursively in case of nested gene sets. To enable transcript-specificity, METRADE was also applied at the splice-isoform level, with annotations retrieved as above.
	The function $\varphi$ maps the expression level of each gene set to a coefficient for the  lower- and upper-limits of the corresponding reaction, and is defined as
	\begin{equation}
	\varphi(\Theta) = \left[\boldsymbol{1}+\gamma \left|log(\Theta)\right|\right]^{\mbox{sgn}(\Theta-\boldsymbol{1})}.
	\label{eq:phi}
	\end{equation}
	Note that the vector notation was adopted, with the convention $0^0=1$ when some element of $\Theta$ is $1$. The sign operator returns a vector of $\pm 1$ (signs of $\Theta-\boldsymbol{1}$). $\gamma$ models the reliability of the gene set expression level as an indicator of the rate of production of the associated enzyme. 
	
	One may argue that gene expression data is not a good proxy for protein abundance and ultimately for flux rates. However, gene expression data is certainly the omic data with better quality and coverage (almost always genome-scale).
	%Quantitative proteomics and RNA-Seq data recently suggested that there is generally high association between mRNA and protein levels in \textit{Escherichia coli} and yeast \cite{guimaraes2014transcript,csardi2015accounting}. A logarithmic map was successfully used for bacterial phenotypic predictions \cite{angione2015predictive}, and is able to capture recent results showing that protein synthesis positively correlates with  mRNA abundance but exhibits lower rate of increase when the mRNA abundance is high \cite{firczuk2013vivo}. 
	Furthermore, in mammals, mRNA levels can be considered the main contributors to the overall protein expression level \cite{li2014system,jovanovic2015dynamic}. Positive correlation between mRNA and protein levels was also recently found in most normal and cancer cell lines \cite{kosti2016cross}.
	We remark that, although GEMsplice natively handles absolute expression values, it can also handle fold-change values directly, by replacing the logarithmic map (\ref{eq:phi}) with $\varphi(\Theta) = \Theta^\gamma$  (see README file in the source code). Furthermore, our method is general and not dependent on a particular proxy. For instance, one may use protein abundances within the same approach proposed here. All simulations were carried out in Matlab R2016b and Octave 4.0.3, with the GLPK solver.

	\paragraph*{\bf Modeling cancer biomarkers in multi-level linear programming.}
	
	The three objectives of the linear program in Equation (\ref{eq:trilevel}) are chosen to model known mechanisms and key players in cancer metabolism. Here we consider: (i) the biomass (growth rate) reaction, (ii) pyruvate kinase (PK) and (iii) lactate dehydrogenase (LDH). (Note however that this is fully customizable in GEMsplice.)
	%Pyruvate kinase is the last step of the glycolytic pathway. Its role is producing ATP and pyruvate from phosphoenolpyruvate (PEP) and ADP.
	%Specifically, the high-energy phosphate in PEP is used to phosphorylate ADP, therefore obtaining ATP and pyruvate. 
	In our pipeline, we minimize the maximum allowed flux rate of PK as the second level of the trilevel linear program, selected through the vector $g$.
	As a third (outer) level of our linear program, selected by the vector $h$, we model the Warburg effect by minimizing the flux through LDH. In fact, a negative flux rate of this reversible reaction models the production of lactate from pyruvate. The first level (inner level), governed by the vector $f$, is assigned to represent the maximization of the flux through the biomass reaction, modeling the observation that cancer cells grow and divide faster than normal cells in order to achieve their main goal, proliferation.

\section{Results}

\subsection*{GEMsplice maps RNA-Seq expression levels and splice isoforms onto a genome-scale model of breast cancer}

\begin{figure}
	\centering
	\includegraphics[width=\columnwidth]{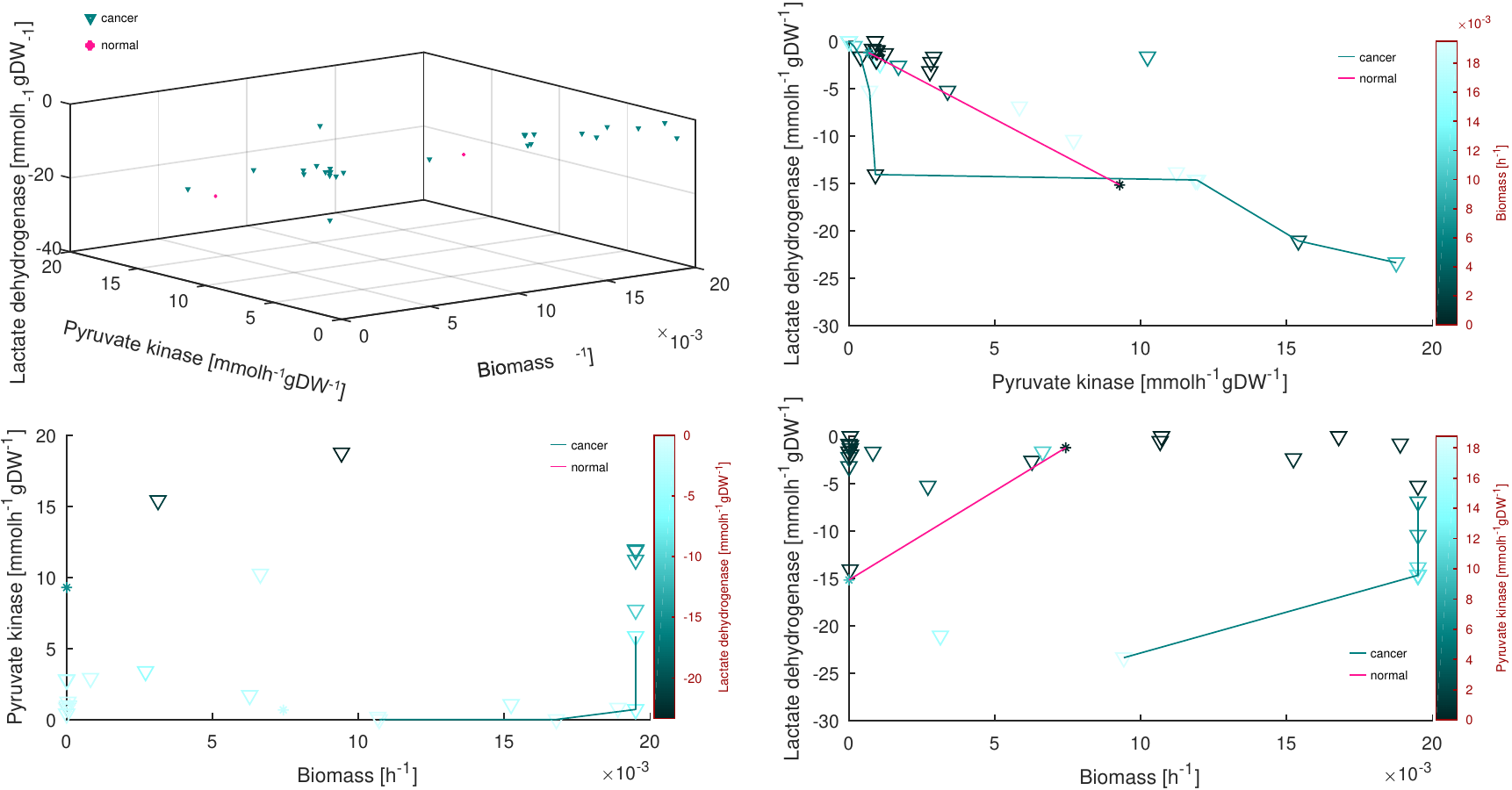} 
	\linespread{1} %force single line spacing
	\vspace{-0.4cm}
	\caption{{\bf The $31$ RNA-Seq profiles representing different breast cancer subsets are mapped to the tridimensional space of biomass, pyruvate kinase and lactate dehydrogenase.} The trilevel linear program (\ref{eq:trilevel}) is solved after constraining the breast cancer metabolic model using each RNA-Seq expression profile from Cancer RNA-Seq Nexus, including splice-isoform expression levels. The bottom left and both right panels show the three projections onto each pair of axes. In each of these, the optimal breast cancer profiles are highlighted and connected for both cancer and normal profiles to identify the trade-off Pareto optimality.}
	\label{fig:biomass_lactate_pyruvate}
\end{figure}

GEMsplice maps RNA-Seq expression profiles to the metabolic model of breast cancer, using for the first time splice isoform annotations. A cohort of $1455$ patients from Cancer RNA-Seq Nexus was taken into account with $31$ different breast cancer subsets, $13$ of which are invasive carcinomas at different stages. 
The subset classification is built from observation of genotype and phenotype, and can for instance refer to a specific type of stage of cancer, a disease state, or a particular cell line. 
%More specifically, each subset is obtained from a group of RNA-Seq samples associated with a specific phenotype or genotype of breast cancer. Depending on the phenotype or genotype observed, a subset can for instance refer to ...
To assess differences in the metabolic response among the different types and stages of breast cancer, we consider the average of RNA-Seq expression levels within the same cancer subset, and we finally map these to the phenotypic space of flux rates in the metabolic model using trilevel optimization (Figure \ref{fig:biomass_lactate_pyruvate}, Figure \ref{sup_fig:biomass_lactate_pyruvate} and File S2; see \textit{Methods} for details on how phenotype predictions are achieved from RNA-Seq data).
GEMsplice correctly predicts breast cancer cells to grow faster than normal healthy cells, while keeping comparable levels of PK activity. Some breast cancer cells are also predicted to use less PK compared to the healthy counterpart, leading to accumulation of intermediates in the glycolytic pathway, therefore fueling the serine biosynthesis pathway \cite{locasale2011phosphoglycerate}. The trade-off between high and low PK flux might also explain the balance between PKM2 (limiting step, controlling PK flux) and PKM1 (promoting OXPHOS in the mitochondrion instead).
The trade-off between LDH and PK in cancer cells is also lower than in normal cells in terms of negative flux rate values, consistent with the widely accepted evidence that cancer cells use the reverse direction of LDH to produce lactate from pyruvate (Warburg effect).
When all the $31$ breast cancer subsets are considered, the bottom right panel shows a weak negative correlation between biomass and negative flux of lactate dehydrogenase (Pearson's $r = -0.34$, $p\mbox{-value} = 7.60 \cdot 10^{-2}$, Spearman's $\rho = -0.37$, $p\mbox{-value} = 5.58 \cdot 10^{-2}$). These predictions are in keeping with the widely accepted fact that the amount of lactate production is positively correlated with tumor growth. More specifically, in breast cancer, a positive correlation has been highlighted between degree of malignancy, degree of mitochondrial structural abnormality \cite{elliott2012mitochondria}, and the most common biomarkers of malignant tumor, e.g.\ intensive use of glycolysis and increased production of lactate \cite{gonzalez2012bio}.

\subsection*{Splice isoform expression-based flux control analysis}

To further analyze the cancer-specific models, and to assess the role of the expression of splice isoforms in the model, we propose a steady-state control analysis based on transcript specificity. This method builds on the technique named \textit{metabolic control analysis} \cite{kacser1995control}, which  considers the relationship between enzyme activity and flux rates. We adapt this analysis to assess the contribution of each splice isoform to the cellular goals.

As described above, flux rates in our model also depend on the expression level of splice isoforms. In each sample, a control analysis on flux rates $v_i$ with respect to these expression levels $x_j$ involves the estimation of the relation between fractional changes in the flux rates and fractional changes in the splice isoform expression. This can be written as a scaled partial derivative of the form
$
C = \frac{x}{v} \frac{\partial v}{\partial x}.
$

We here adopt an adjusted control coefficient by shrinking the sample-specific denominator $v$ towards its average $\bar{v}$ across all the samples.  As in the standard deviation correction for differential expression tests \cite{tusher2001significance}, our correction prevents gross overestimation of control coefficients. Without this correction, ``false positives'' would arise when $v_i(x_j)$ is a very small value, irrespective of the numerator. More formally, to approximate this partial derivative, we separately take into account each splice isoform $j=1,...,N$ included in the breast cancer metabolic model, and we evaluate positive and a negative intermediate control coefficients $C_{ij}^{+}$ and $C_{ij}^{-}$, defined as
\begin{equation}
%\begin{split}
C_{ij}^{\pm} = \frac{v_i(x_j \pm \delta) - v_i(x_j)}{(v_i(x_j)+\bar{v_i})/2} \bigg/ \frac{\delta}{x_j}, \ i=1,...,M, \ j=1,...,N,\\
%\end{split}
\end{equation}
where $\delta$ is small enough so that the ratio approximates the derivative; $v_i$ is the flux of interest (in our analysis, we focus on the three objectives of the linear program, namely biomass, PK and LDH).
%Note that the use of fractional changes allows us to evaluate the effect of each splice isoform's expression level independent of the units adopted to measure it. 
Then, for each flux and splice isoform of interest, we finally calculate the overall \textit{flux control coefficient} $C_{ij}$ as the maximum variation caused by a positive or negative perturbation of the isoform expression level: 
\begin{equation}
C_{ij} = \text{max} (\left|C_{ij}^{+}\right|,\left|C_{ij}^{-}\right|).
\end{equation}

These transcript- and cancer-specific coefficients evaluate the relative steady-state change in three pivotal flux rates in breast cancer cells, with respect to the relative change in the expression level of the transcript (see File S3). For our simulations, we set the denominator $x_j = 1$ to avoid proportionality between the final control coefficient and the expression value itself, and $\delta=10^{-3}$. In general, the smaller the value chosen for $\delta$, the more effective $C$ in approximating the scaled derivative of $v_i$. Note that calculating a flux control coefficient for each splice isoform does not explicitly identify controlling transcripts, but rather provides an effective way to quantify their influence on the key flux rates in breast cancer cells.

The $10$ most influential transcripts were selected independently for biomass, PK and LDH. The union of these three sets of ten transcripts, composed of $14$ transcripts, was chosen to perform further enrichment analysis through PANTHER \cite{mi2016panther}. A functional classification of each transcript was obtained using the ``protein class'' ontology, which is adapted from the PANTHER/X molecular function ontology and also includes Gene Ontology (GO) annotations. Interestingly, the selected transcripts are highly enriched for transmembrane transport and respiratory electron transport chain.

As shown in Figure \ref{fig:bubble_pathways}a, the most effective transcript at controlling the value of biomass and LDH with a less stringent control on PK is ENST00000330775, a transcript of glucose-6-phosphate translocase (G6PT). In brain cancer,  G6PT is a key player in transducing intracellular signaling events; modulating its expression has been proposed as an anti-cancer strategy \cite{belkaid2006silencing}. Furthermore, our results suggest that PK and LDH are maximally and simultaneously controlled by ENST00000591899 and ENST00000378667, transcripts of ubiquinol-cytochrome c reductase, whose amplification was suggested to correlate with more aggressive breast cancer \cite{ohashi2004ubiquinol}. ENST00000507754 and ENST00000327772, transcripts of Complex I (whose activity is known to regulate breast cancer progression \cite{santidrian2013mitochondrial}), can control both PK and LDH without disrupting the growth rate.  Although further experimental investigation is needed on these key transcripts, our genome-scale method may suggest for the first time transcript-specific targets for anti-cancer strategies.

%\begin{figure}
%\centering
%\includegraphics[width=0.5\columnwidth]{figures/mca.pdf}
%\linespread{1} %force single line spacing
%\caption{{\bf Transcript-based metabolic flux analysis.}  The analysis is applied to the set of $16$ transcripts obtained as a union of the three sets of ten most influential transcripts with respect to biomass, pyruvate kinase and lactate dehydrogenase flux rate. Error bars represent the standard error of the mean $\mbox{\it SE} = \sigma/\sqrt{n}$, where $\sigma$ is the standard deviation of the measured effect, computed across all breast cancer subsets, and $n$ represents the number of breast cancer subsets.}
%\label{fig:mca}
%\end{figure}

\subsection*{Pathway-based flux analysis}

A pathway-based perspective has been often taken in genome-scale models with the aim of investigating sensitivity analysis \cite{kent2013can,costanza2012robust,conway2016iterative}, and coupled with Bayesian techniques to detect pathway crosstalks and temporal activation profiles \cite{angione2015hybrid}.
To assess the variation in the average flux of each pathway with respect to the unconstrained breast cancer model, we here compute a normalized average pathway flux
\begin{equation}
d_i = \left({\bar{w}^{(i)} - w_U^{(i)}}\right)/{w_U^{(i)}}, \quad i=1,...,P,
\end{equation}
where $\bar{w}^{(i)}$ is the average flux in the $i$th pathway across the different cancer subsets, while $w_U^{(i)}$ indicates the flux of the $i$th pathway in the unconstrained breast cancer model. To account for the tolerance of the linear solver, average pathway fluxes of less than $10^{-10}$ were assumed to be zero. Pathways, defined using the KEGG LIGAND database, were inherited from Recon1.

\begin{figure*}
	\centering
	\begin{tabular}{ccc}
		\includegraphics[width=0.34\textwidth]{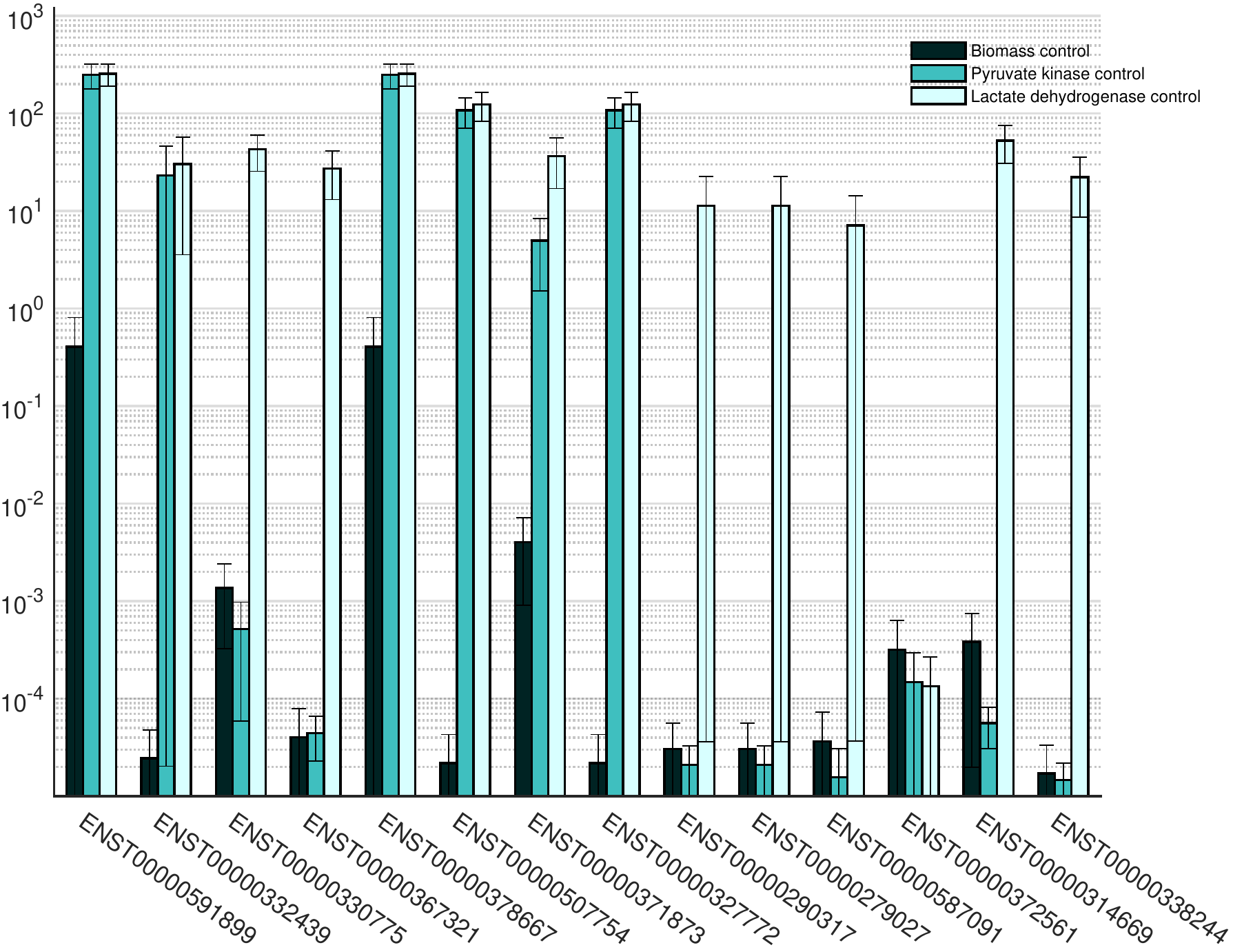} &
		\hspace{0.4cm}\includegraphics[width=0.29\textwidth]{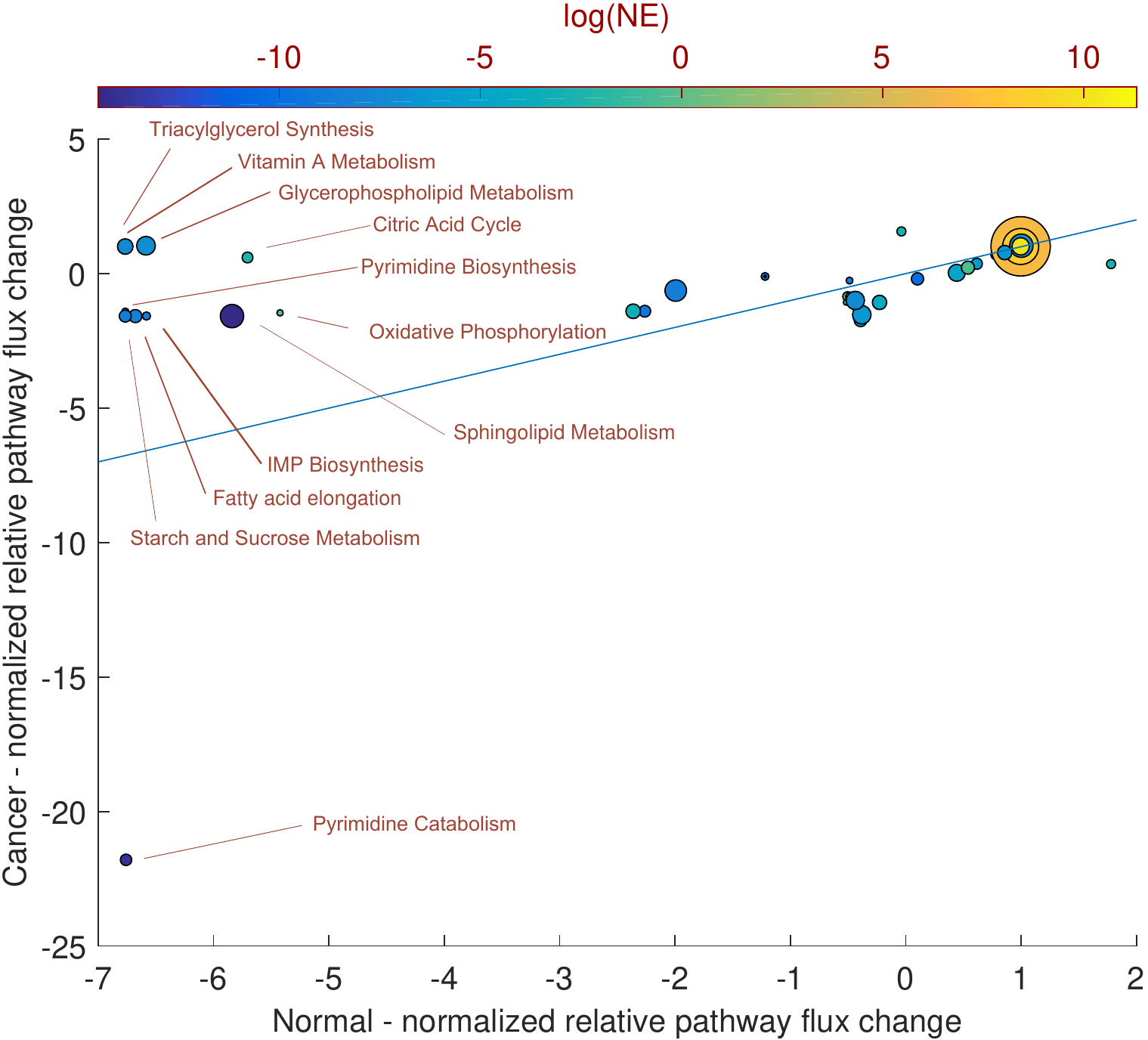} & \hspace{0.6cm}\includegraphics[width=0.28\textwidth]{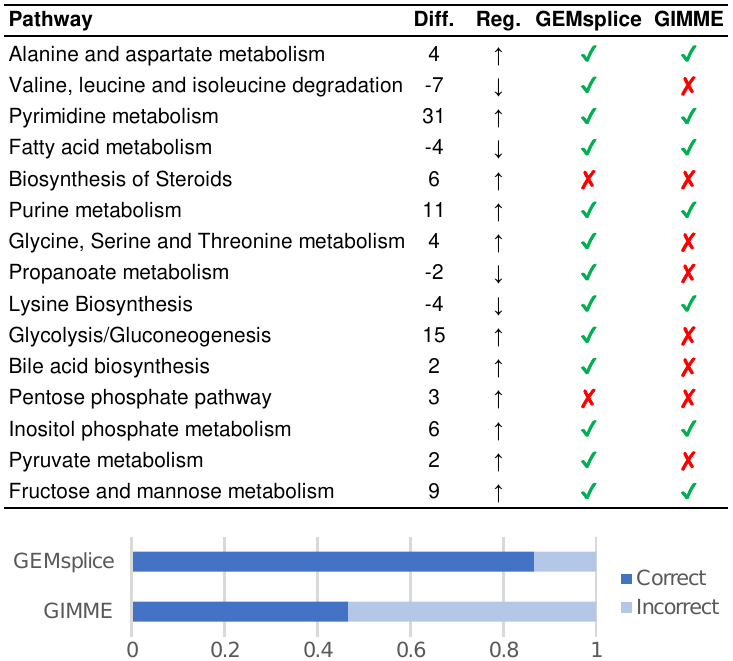}\\
		(a) & (b) & (c)
	\end{tabular}
	\linespread{1} %force single line spacing
	\vspace{-0.3cm}
	\caption{{\bf Transcript-based metabolic flux and pathway analysis.}  (a) The control analysis is applied to the set of $14$ transcripts obtained as a union of the three sets of ten most influential transcripts with respect to biomass, pyruvate kinase and lactate dehydrogenase flux rate. Error bars represent the standard error of the mean $\mbox{\it SE} = \sigma/\sqrt{n}$, where $\sigma$ is the standard deviation of the measured effect, computed across all breast cancer subsets, and $n$ represents the number of breast cancer subsets. 
		(b) The indicator $d$ plotted for each pathway in breast cancer versus normal breast cells (adjacent to tumor cells). The size of each point represents the size of the corresponding pathway, quantified as the number of reactions belonging to it. 
		The $y=x$ line is shown on the plots to highlight outliers. By definition of $d$, if a pathway lies outside the line, its perturbation in cancer samples is different from its perturbation in healthy samples. (Both perturbations are computed with respect to the default breast cell model run without omic-derived constraints.)
		Colors are given according to a normalized error of the computed mean flux in each pathway, namely $\mbox{\it NE} = \sigma/\sqrt{p}$, where $\sigma$ is the standard deviation of the flux rate in the pathway, computed across all breast cancer subsets, and $p$ represents the size of the pathway. Note that the average flux in a pathway is positive or negative depending on the direction in which its reactions take place.
		(c) GEMsplice correctly predicts 13 out of 15 pathway increased/decreased activity (difference ``Diff'' between upregulated and downregulated reactions), while GIMME only predicts 7 out of 15. More accurate predictions are achieved for key cancer biomarkers: glycolysis, glycine, serine, and threonine metabolism, and valine, leucine and isoleucine degradation.}
	\label{fig:bubble_pathways}
\end{figure*}

Figure \ref{fig:bubble_pathways}b shows the indicator $d$ plotted for each pathway when computed across breast cancer cells and normal cells (results for invasive and unlabeled breast cells are shown in Figure \ref{sup_fig:bubble_pathways}). 
The least-square linear regression  reveals a positive pathway flux correlation (Pearson's $r = 0.43$, $p\mbox{-value} = 1.82 \cdot 10^{-3}$, Spearman's $\rho = 0.80$, $p\mbox{-value} = 2.99 \cdot 10^{-12}$). The eleven pathways indicated in the figure were detected as outliers, with significantly different behavior between cancer and normal breast cells.

To support our predictions, literature-based evidence of breast cancer alterations in such pathways was aggregated.
Alanine and aspartate are byproducts of amino acid and glutamine fermentation, an important source for protein synthesis, especially when the cell lacks oxygen.
Aspartate is also a main contributor to nucleotide and protein synthesis; the malate/aspartate shuttle pathway translocates electrons for the mitochondrial electron transport chain to produce ATP \cite{pecqueur2013targeting}.
Perturbations in the citric acid cycle (TCA) were largely expected because ot the known ``Warburg'' reduction of glucose metabolism through the mitochondrion.  Likewise, the prediction of altered fatty acid elongation metabolism is in keeping with recent studies reporting elongation of fatty acids as a marker in breast cancer \cite{feng2016elovl6}. 

GEMsplice also correctly predicts that pyrimidine biosynthesis pathways are downregulated in normal cells when compared to breast cancer cells \cite{sigoillot2004breakdown}. Vitamin A and carbohydrate altered metabolism are also associated with breast cancer \cite{doldo2015vitamin,nagarajan2016oncogene}. 
In breast cancer cells, Inosine monophosphate (IMP) dehydrogenase inhibitors cause growth reduction and phenotypic alterations \cite{sidi1988growth}. The prediction of starch and sucrose metabolism confirms previous genome-wide association studies, which reported significant relation between altered starch and sucrose pathway and the risk of developing ER-negative breast cancer \cite{li2010genome}.
Sphingolipids are responsible for oestrogen signaling and can control differentiation and proliferation of breast cancer cells \cite{sukocheva2014role}. The glycerophospholipid pathway, also flagged as biomarker by GEMsplice, was previously found altered in the metabolic phenotype of breast \cite{cadenas2012glycerophospholipid}. Finally, triacylglycerol was found to be a biomarker and a prognostic predictor of poor clinical outcome in triple negative breast cancer \cite{dai2016pretreatment}.
\subsection*{Comparison with gene-level approaches} 

We here compare the results obtained by GEMsplice with GIMME \cite{becker2008context}, where expression levels are mapped to the model with gene-level resolution only. Present/absent calls for GIMME were obtained by applying zFPKM normalization \cite{hart2013finding}.
To compare the pathway activity predicted by GEMsplice and GIMME in cancer conditions, we computed the average absolute value of flux rates in all nonzero biomass samples. 
As a benchmark for comparison and validation, we used the set of metabolic pathways found to be significantly upregulated and downregulated in unfavorable breast cancer conditions \cite{schramm2010analyzing}. Pathways with equal number of upregulated and downregulated reactions were excluded from the analysis.  

GEMsplice correctly predicts $13$ out of $15$ pathway overexpression/underexpression patterns, while GIMME only predicts $7$ out of $15$ (Figure \ref{fig:bubble_pathways}c, and File S5). Three key pathways widely accepted to be dysregulated in breast cancer are correctly identified by GEMsplice, but incorrectly by GIMME: (i) glycolysis, highly upregulated in cancer, shows 3.44 cancer/normal fold change in GEMsplice, while 0.55 in GIMME; (ii) glycine, serine, and threonine metabolism, upregulated in cancer: 2.54 fold change in GEMsplice, 0.41 in GIMME; (iii) valine, leucine and isoleucine degradation, downregulated in cancer, 0.30 fold change in GEMsplice, 2.85 in GIMME.

The exact flux rates predicted for lactate dehydrogenase activity were also evaluated against the widely-accepted cancer ``Warburg'' metabotype. Ten cancer samples are incorrectly predicted by GIMME with zero or negligible ($<10^{-12}$) ``Warburg'' LDH flux (therefore not switching to aerobic glycolysis), while only three by GEMsplice. 
Furthermore, all non-malignant and normal cells with nonzero growth are correctly predicted by GEMsplice with lowest activity of LDH, while two out of four nonzero-growth normal or non-malignant tissues are incorrectly predicted by GIMME with a typical cancer metabotype, consisting of very high LDH activity.
These examples suggest that integrating splice isoform information with metabolic models improves the characterization of breast cancer metabolism. With the current fast-paced improvements in speed and costs of omic profiling, and therefore with increasing availability of isoform-level model annotations and RNA-Seq data, we expect such specific differences to increase even further.

%\section{Discussion}

%%%%%%%%%%%%%%%%%%%%%%%%%%%%%%%%%%%%%%%%%%%%%%%%%%%%%%%%%%%%%%%%%%%%%%%%%%%%%%%%%%%%%
%
%     please remove the " % " symbol from \centerline{\includegraphics{fig01.eps}}
%     as it may ignore the figures.
%
%%%%%%%%%%%%%%%%%%%%%%%%%%%%%%%%%%%%%%%%%%%%%%%%%%%%%%%%%%%%%%%%%%%%%%%%%%%%%%%%%%%%%%

\section{Conclusion}

While the rate of acquisition of omics data is rapidly increasing, analyzing and extracting information through computational tools arguably remains the main bottleneck in biology. Most studies focus on statistical analysis on gene expression values to evaluate how they vary across different samples. However, modeling how the gene expression alterations change metabolic processes at genome scale provides greater understanding of the phenotypic outcome with respect to studies involving only transcriptomic data \cite{angione2016multiplex}. 
Given that the last decade in cancer research has repeatedly shown that cancer is a complex disease that cannot be studied by narrowing it down to a single gene or enzyme, a genome-wide metabolic approach seems therefore the right direction to assess and predict the phenotype of a cancer cell. 

We here proposed GEMsplice, a method for linking gene expression and splice isoform data to genome-scale metabolic models. 
GEMsplice is the first attempt at solving one of the main issues of metabolic modeling: as outlined in recent reviews \cite{geng2017silico,pfau2016towards,yizhak2015modeling,ryu2015reconstruction}, current methods only allow integration of omics data up to the gene level, but not with splice-isoform resolution.

The idea is that every single profile can be used to create a profile-specific model of metabolism that includes splice-isoform annotations. This integrated model can be readily used to predict the flux rate of any biochemical reaction included in the metabolic model, in a range of cancer subtypes. Individual profiles, patients or cells related to the same cancer can be mapped onto this model, for instance to cross-compare their metabolic behavior within the same cancer type, instead of cross-comparing cells using transcriptomics data only. 

Although it allows mapping cancer-specific transcript expression levels onto any metabolic network, GEMsplice comes with limitations and room for improvement. For instance, if reaction-specific information is available, the map from genes to reaction flux bounds could be chosen in a reaction-specific manner and tailored to specific environmental or growth conditions. Likewise, post-transcriptional regulation of expression levels could be readily included in the model if available. On the other hand, inconsistencies can be used to improve the model and the associated biological knowledge on metabolic enzymes. For instance, incorrect predictions of specific flux rates may shed light on where post-transcriptional regulation take place \cite{markert2015mathematical}.
A further development of cancer metabolic studies may involve a model that takes into account interactions between cancer cells and their environment, e.g. interactions with supporting cells. This approach will likely capture features that cannot be observed with models of single cancer cells. 

Breast cancer is manifested through a variety of effects that cannot be reduced to a single feature. The difference in behavior of different cancer cells shows that reconstructing a generic model of a cancer cell is not a viable approach \cite{ghaffari2015identifying}. Models should therefore always be created in a tissue- and stage-specific fashion. We anticipate that GEMsplice will allow for the first time the generation of such models harnessing the full potential of RNA-Seq, and will facilitate in silico combinatorial experiments with RNA-Seq data. In fact, such omic-informed models can now effectively integrate information on the expression level of splice isoforms, to date largely ignored.

\section*{Acknowledgements}
The author would like to thank Dr Jim Liu for discussions on Cancer RNA-Seq Nexus, and Dr Syed Haider for discussions on splice isoforms.

\begin{scriptsize}
\bibliographystyle{unsrt}
\bibliography{biblio_gemsplice}
\end{scriptsize}

%% Authors are advised to submit their bibtex database files. They are
%% requested to list a bibtex style file in the manuscript if they do
%% not want to use elsarticle-num.bst.

%% References without bibTeX database:

% \begin{thebibliography}{00}

%% \bibitem must have the following form:
%%   \bibitem{key}...
%%

% \bibitem{}

% \end{thebibliography}

\end{document}

\maketitle              % typeset the title of the contribution

\abstract{\textbf{Motivation:} Despite being often perceived as the main contributors to cell fate and physiology, genes alone cannot predict cellular phenotype. During the process of gene expression, $95\%$ of human genes can code for multiple proteins due to alternative splicing.  While most splice variants of a gene carry the same function, variants within some key genes can have remarkably different roles.
To bridge the gap between genotype and phenotype, condition- and tissue-specific models of metabolism have been constructed. However, current metabolic models only include information at the gene level. Consequently, as recently acknowledged by the scientific community, common situations where changes in splice-isoform expression levels alter the metabolic outcome cannot be modeled. \\
\textbf{Results:} We here propose GEMsplice, the first method for the incorporation of splice-isoform expression data into genome-scale metabolic models. Using GEMsplice, we make full use of RNA-Seq quantitative expression profiles to predict, for the first time, the effects of splice isoform-level changes in the metabolism of $1455$ patients with $31$ different breast cancer types. We validate GEMsplice by generating cancer-versus-normal predictions on metabolic pathways, and by comparing with gene-level approaches and available literature on pathways affected by breast cancer.
GEMsplice is freely available for academic use at \url{https://github.com/GEMsplice/GEMsplice_code}.
Compared to state-of-the-art methods, we anticipate that GEMsplice will enable for the first time computational analyses at transcript level with splice-isoform resolution.
}

\subsection*{Flux Balance Analysis (FBA)}
To generate cancer-specific models from Cancer RNA-Seq Nexus, we propose an extended flux balance analysis (FBA) framework.
FBA has been successfully applied to predict growth rate and other metabolic observables in cells \cite{palsson2015systems}. Starting from the stoichiometry matrix $S$ of all known metabolic reactions in a cell, and given an unknown vector $v$ of reaction flux rates, FBA models the steady-state of a cell through the condition $Sv=0$. Further constraints are added as a lower- and upper-limit for the flux rate of each reaction. Through linear programming, a combination of flux rates is chosen as an objective to be maximized or minimized. The underdetermined linear system $Sv=0$ is then solved and a distribution of flux rates is determined according to the objectives chosen and the constraints added on the flux rates.
Standard FBA is extended through METRADE \cite{angione2015predictive} to achieve multi-omic integration and multi-level linear programming, as detailed in the main manuscript.

\subsection*{GEMsplice pipeline}
To build GEMsplice, we performed manual curation of the model using data associated with each gene in Recon1, retrieving transcript level information and all available identifiers published as supplementary material but not included in the model. We then mapped transcript identifiers in the Recon1 SBML file to transcript IDs used by databases, using BioMart \cite{smedley2015biomart} and Ensembl \cite{yates2016ensembl}. We finally used these IDs within the breast cancer metabolic model by Jerby et al.~\cite{jerby2012metabolic}. Using a RNA-Seq dataset and a method based on constraint-based programming, we were able to map transcript expression levels, including splice isoforms, onto the breast cancer model in order to generate $31$ cancer-specific metabolic models.
Cancer RNA-Seq Nexus was used as a RNA-Seq dataset \cite{li2016cancer}, including data from TCGA \cite{cancer2012comprehensive}, GEO \cite{barrett2013ncbi} and SRA \cite{kodama2012sequence}.

\subsection*{Analysis of invasive versus unlabeled cancer cells}

\begin{figure}
	\centering
	\begin{tabular}{cc}
		\includegraphics[width=0.5\columnwidth]{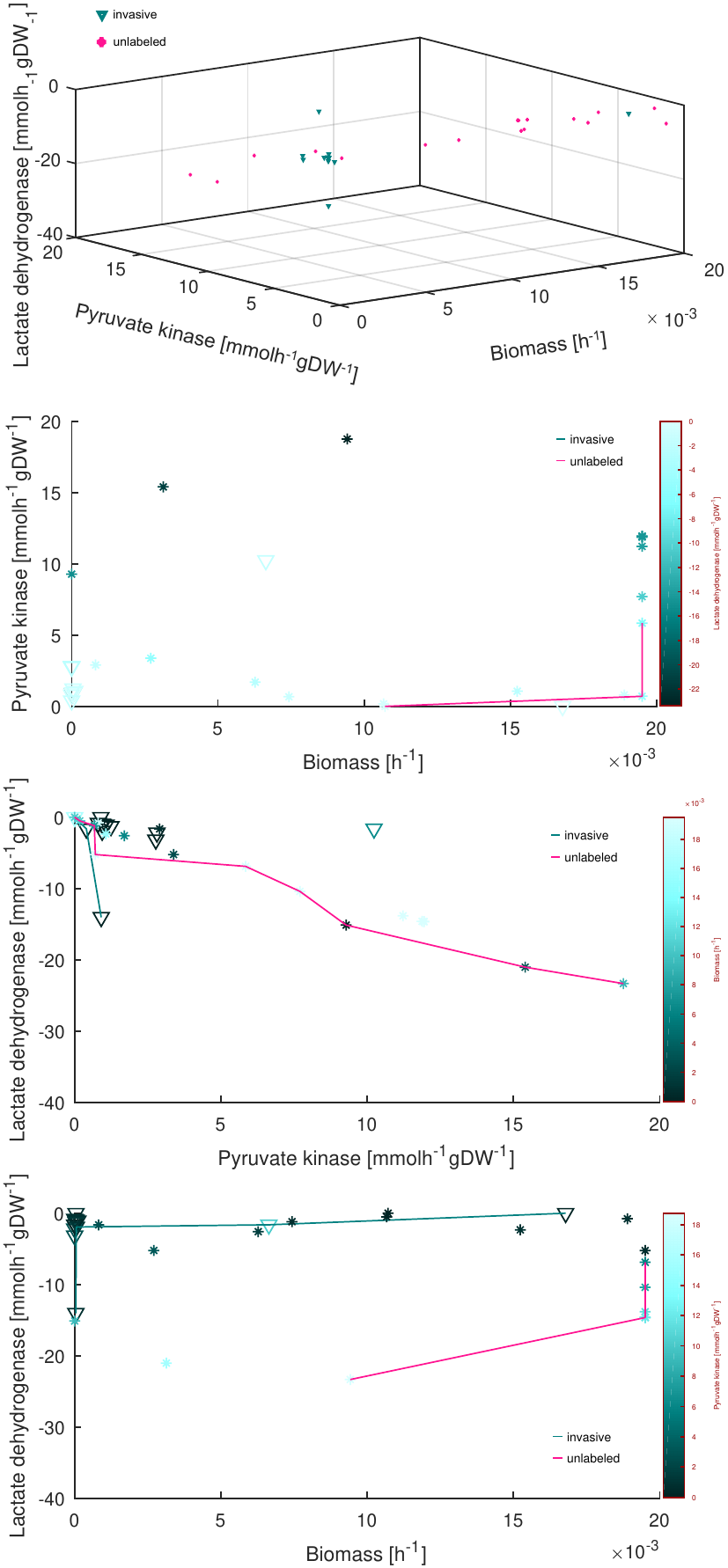} 
	\end{tabular}
	%\vspace{-0.3cm}
	\linespread{1} %force single line spacing
	%\vspace{-0.5cm}
	\caption{{\bf The $31$ RNA-Seq profiles representing different breast cancer subsets are mapped to the tridimensional space of biomass, pyruvate kinase and lactate dehydrogenase.} The trilevel linear program (\ref{eq:trilevel}) is solved after constraining the breast cancer metabolic model using each RNA-Seq expression profile from the Cancer RNA-Seq Nexus dataset, including splice-isoform expression levels. The bottom panels show the three projections onto each pair of axes. In each of these two-dimensional objective spaces, the optimal breast cancer cells (in terms of trade-off Pareto optimality) are found and connected by a line for invasive and unlabeled cells. Unlabeled cells are normal cells or cells that have not been detected as invasive.}
	\label{sup_fig:biomass_lactate_pyruvate}
\end{figure}

Overall, unlabeled cancer cells, defined as the normal cells and the cells that have not been clearly detected as invasive, are predicted to produce more biomass than invasive cells. The trade-off frontiers in Figure \ref{sup_fig:biomass_lactate_pyruvate} suggest that invasive cancer cells outperform unlabeled cells in carrying a positive flux of pyruvate kinase while ensuring a large flux for lactate dehydrogenase. More specifically, when comparable flux rates of pyruvate kinase are considered, the invasive breast cancer cells considered in our study are able to achieve a higher flux for the reverse reaction lactate dehydrogenase, which in turn represents an increased Warburg effect, i.e., an increased production of lactate from pyruvate. This result supports the hypothesis that aggressive phenotype and worse clinical outcome are positively correlated with Warburg effect \cite{gatenby2004cancers}. 

\begin{figure}
	\centering
	\includegraphics[width=0.5\columnwidth]{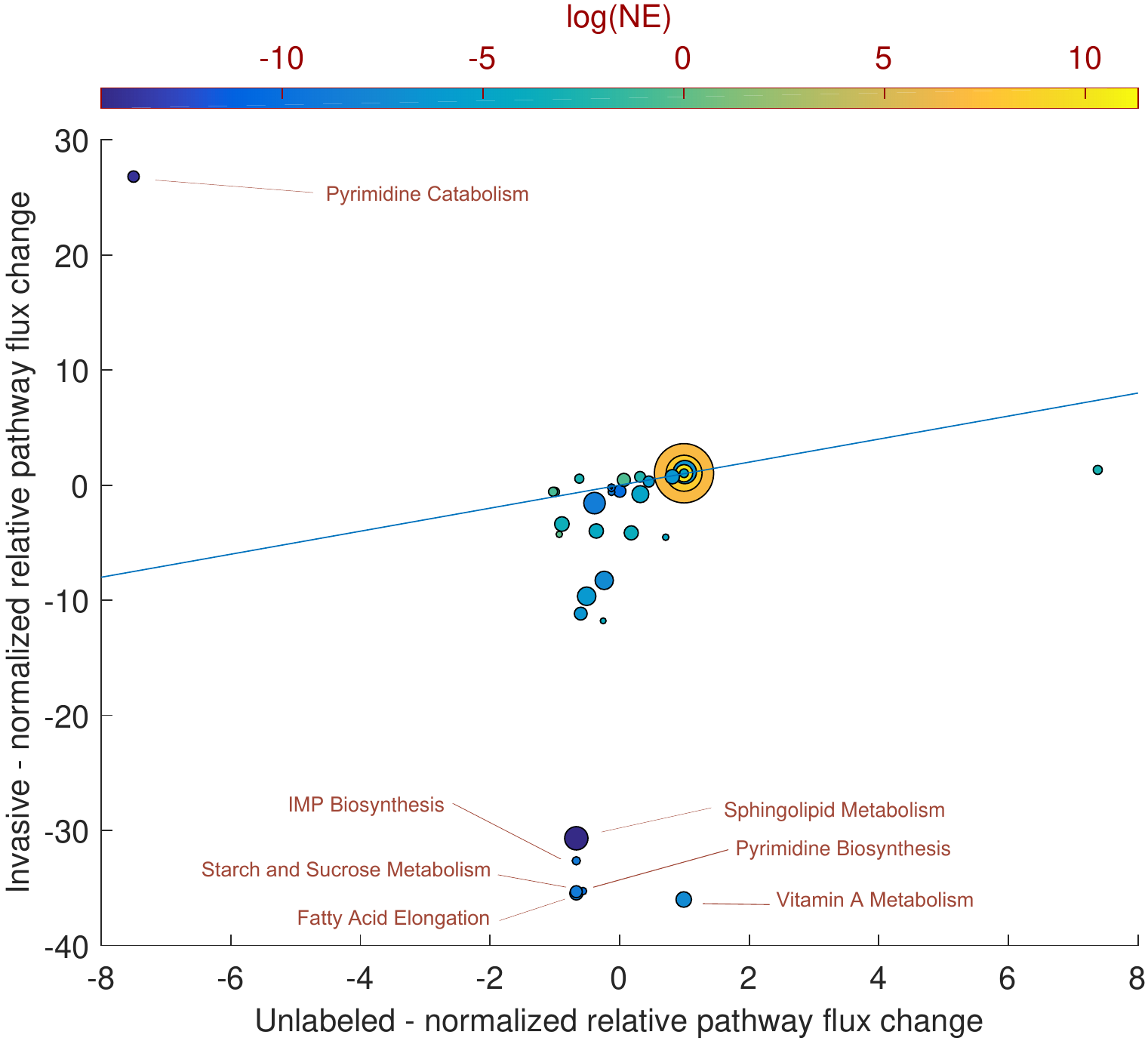}
	\caption{{\bf Transcript-based pathway analysis.} Indicator $d$ plotted for invasive breast cancer versus unlabeled breast cells. The size of each point represent the number of reactions in the pathway. Colors are given according to a normalized error of the computed mean flux in each pathway, namely $\mbox{\it NE} = \sigma/\sqrt{p}$, where $\sigma$ is the standard deviation of the flux rate in the pathway, computed across all breast cancer subsets, and $p$ represents the size of the pathway. The $y=x$ line is shown on the plots to highlight outliers. By definition of $d$, if a pathway lies outside the line, its perturbation in cancer samples is different from its perturbation in healthy samples (both perturbations are computed with respect to the default breast cell model run without omic-derived constraints).}
	\label{sup_fig:bubble_pathways}
\end{figure}

The  pathways detected as outliers, with significantly different behavior between invasive and unlabeled breast cell, are reported in Figure \ref{sup_fig:bubble_pathways}.The plot also shows overall positive correlation between invasive and unlabeled perturbations (Pearson's $r = 0.22$, $p\mbox{-value} = 1.17 \cdot 10^{-1}$, Spearman's $\rho = 0.63$, $p\mbox{-value} = 3.87 \cdot 10^{-7}$). This suggests that our method behaves consistently across different cancer subsets, but also highlights pathways whose behavior is significantly changed in the two cases.

\begin{scriptsize}
\bibliographystyle{unsrt}
\bibliography{biblio_gemsplice}
\end{scriptsize}

%% Authors are advised to submit their bibtex database files. They are
%% requested to list a bibtex style file in the manuscript if they do
%% not want to use elsarticle-num.bst.

%% References without bibTeX database:

% \begin{thebibliography}{00}

%% \bibitem must have the following form:
%%   \bibitem{key}...
%%

% \bibitem{}

% \end{thebibliography}